# Ethics in the digital era

David Pastor-Escuredo[1,2]

**Abstract.** Ethics is an ancient matter for human kind, from the origin of civilizations ethics have been related with the most relevant human concerns and determined cultures. Ethics was initially related to religion, politics and philosophy to then be fragmented into specific communities of practice. The undergoing digital revolution enabled by Artificial Intelligence and Data are bringing ethical wicked problems in the social application of these technologies. However, a broader perspective is also necessary. We now face global and highly dynamics challenges that affect groups and individuals, specially those that are most vulnerable. Individual-oriented ethics are no longer sufficient, the new ethic has to consider the several scales in which the current complex society is organized and the interconnections between different systems. Ethics should also give a response to the systemic changes in behavior produced by external factors and threats. Furthermore, AI and digital technologies are global and make us more connected and smart but also more homogeneous, predictable and ultimately controllable. Ethic must take a stand to preserve and keep promoting individuals rights and uniqueness and cultural heterogeneity. Digital technologies have to the foundation for new models of society and help ensure ethical individual and collective values. For these reasons science has to be at the core of the new ethic as it helps understand the complex world. Finally, AI has advanced through the ambition to humanize matter, so we should expect ethics to give a response to the future status of machines and their interactions with humans.

## 1 INTRODUCTION

Ethics has always been a fundamental matter for humanity. Since the first civilizations ethical principles guided human behavior (Kramer, 1981). The relationship with divinities and death configured the ethics in the ancient cultures such as Egypt or Mesopotamia (Masson-Oursel, 1938) (OPPENHEIM, 1968). Greek cities brought a new political and collective dimension to ethics along with the first philosophical systems where ethics was central (Hackforth, 1972) (Ameriks and Clarke, 2000). Greek culture was probably the first in having a globalization impetus trying to bridge western and eastern civilizations that was inherited and adapted by romans (Barceló, 2001). However, the later political fragmentation and spread of different religious traditions in Europe and Asia configured an heterogeneous landscape of ethics that characterized cultural and social development (Delort, 1969).

[1] Center Innovation and Technology for Development, Technical University Madrid, email: david.pastor@upm.es
[2] LifeD Lab, Madrid, Spain

The irruption of the Enlightenment produced a disruptive shift in the conception of ethics. A human-centered vision prompted by the maturity of humanity, philosophical reasoning and science changed how ethics was justified and formulated (Kant, 2002). However, the aspiration for a common human ethical system based on human reasoning did not last. Post-modernity, capitalism and economic globalization led to a mosaic of metacultures and social trends spread around the globe with a strong sense of individuality (Sartre et al., 2004). As a result, philosophical paradigms and therefore ethics became highly fragmented and enclosed within communities of practice and finely delimited areas of knowledge. Ethics was forced to establish their own limits (Muguerza, 2004).

Technological development has become one of the main drivers of socio-economic configuration of the world map. The Internet enabled hyper-connectivity and real-time flows of information with strong implications in communication and financial systems. The awareness of the power of the data generated by the society through information systems and portable devices brought the new data revolution. Thanks to data, machines have been able to learn and produce more valuable and actionable insights. Thus, Artificial Intelligence is now reaching its promised potential (Simon, 1969) (Stone et al., 2016). It has been claimed that AI and data are the core of the emergent industrial revolution, however, this revolution may have deeper implications beyond socio-economic dimensions. In this light, ethics is gaining attention and traction because the way machines may behave and decide bring back questions about how humans should behave and decide. A very important feature of technological development and implications has been that its pace is way faster than the reasoning about it including ethics.

In the recent years we have witnessed an international effort to address global challenges. Starting with the Millennium Development Goals, countries have reached a global consensus in terms of Sustainable Development Goals that configure the international development agenda (Sachs, 2012) (Griggs et al., 2013). Poverty, gender, inequality, vulnerability and how we live in our planet have become central issues for humankind and governments worldwide. The long-term development goals are also challenged by crisis of different kind (natural disasters, economic crisis, pandemics, etc) that may produce systemic changes, potentially irreversible, in the society. Social dynamics are more than ever a matter of ethics.

Here, the key aspects for ethics in this digital era are presented. The challenges that emerge are also a new opportunity to revitalize ethics in a more global sense so that we succeed in building a better humanity. Therefore, it is necessary to provide ethic with contents to fulfill this mission: revise individual values, develop collective values, learn to live in a complex society and create ethics with a sound scientific ground.

## 2 GLOBAL CHALLENGES NEEDS A SYSTEMIC APPROACH

The digitally-enabled hyper-connected world brings important benefits for the society. Remarkably, knowledge can be more openly distributed and exploited. Artificial Intelligence and Data are the new elements of an emergent socio-economic revolution, but this revolution goes beyond industrial. In a global scale, the interconnections between different global subsystems, the real-time information and the social response to information (and misinformation), events and crisis makes the world a tangled network of complex processes. At the individual level, the use of digital platforms, apps and tools facilitate people's life in many aspects and generate massive amounts of behavioral data with very high-value provided AI-based tools that analyze it. Thus, the local and the global scales are more interconnected than ever.

The structure of the society is multi-scale and multi-layered, therefore, a mere individual approach to ethics is now doubtfully appropriate, the collective dimension has become critical. As the digital revolution is a global driver, we need ethics that addresses digitalization as a global phenomenon and how it affects people worldwide. To this matter, the philosophical tradition provides reflections on the inter-individual foundation of ethics (Kelly et al., 1994). We need to build on the philosophical tradition and deep scientific knowledge of systems for ethics with a global scope.

We are also challenged by threats of different nature (natural hazards, economic crisis, pandemics, etc) whose multi-dimensional impact scales up in time and space and also affects more to those that are vulnerable. Crisis have important implications that ethics cannot overlook: systemic risk is high and many societies are mostly fragile (Taleb, 2012). Risk are often non evident, many social processes and events are characterized by what is known as fat-tail distributions where specific unlikely events have a large impact on the system ("Black swans"). Therefore, they require a different and systemic approach to risk management that involves individual and collective actions and behavioral changes (Norman et al.). But the consequences scale up even more as we are digitally interconnected, there is a net that propagates impact even when it is negative. Digital media play a key role in configuring the narratives that outlast. Furthermore, generations may develop their cognitive system under stress influencing their perception and values (Varela et al., 2016). We need ethics that account for the deep changes produced by crisis and specially the ones more critically exposed.

However, there is a counterpart, this is the serious risk of the core individual rights and values and cultural heterogeneity. The technology more and more influences our daily behavior, as global digital platforms have a large penetration in many parts of the world, our routines can become more homogeneous, predictable and ultimately controllable. Thus, beyond individual privacy, there are additional risks regarding global misinformation (e.g. Deep Fakes), cognitive overflow, thought control (e.g. micro-targeting), sharper inequalities (e.g. digital gap), aggressive surveillance, control lobbies or the commonly known algorithm biases. For instance, data-driven policies and surveillance can be tighten during emergencies and outlast, potentially reducing population rights. Ethics must address these issues and keep promoting individuals, collective well-being and cultural heterogeneity at the same time that faces global challenges.

In summary, in a future society based on digital technologies and AI, the technological machinery has to be the substrate on which the model of society, the ethical values and the vision of humanity (as individuals and collectively) has to be built upon. This means that the digital technologies should help ensure the model and vision works in the same way now we have systems to ensure rights or security. Ethic has to be proactive in the design and foundation of these systems that likely need to have a global scope.

## 3 ETHICS AND ARTIFICIAL INTELLIGENCE

As rapidly overviewed, ethics have a long tradition in philosophical frameworks. Now, ethics has to adopt concepts and language from technologists and develop shared knowledge and language to bring this tradition into the global digitalization, and this has to be done fast. Ethical principles such as liberty, equity or solidarity must be inserted into digitalization at the very core of its whole ecosystem (Jobin et al., 2019) (Luengo-Oroz, 2019).

The digital revolution has been clearly led by the industry and the global digital platforms. Data and AI are being now used to make better decisions, optimize processes, customize services, predict patterns and understand society. As it could be expected, the objectives of companies may not be fully aligned with public interests. Even when social responsibility and impact are becoming a key issues for investors and slowly appearing in regulatory systems, ethics should reflect on a framing to evaluate the uses and impact of AI and digitalization in the terms introduced above. Sustainable Development Goals have proved to be a suitable framework to catalyze and promote uses of Data and AI to address global challenges (Vinuesa et al., 2020). Ethics has a great opportunity to articulate SDGs with individuals, groups and organizations that conform society.

Most of the current ethical questions on AI have appeared due to wicked problems related to the application of AI to real world applications. From automated diagnosis to self-driving cars, ethical issues naturally arise. These issues are now mainly solved by engineers and private sector companies. Algorithm development must be openly discussed to ensure functioning according to ethical guidelines and frameworks. This implies technical, organizational and theoretical challenges that the society is implicitly demanding to improve explainability and interpretability of AI (Pastor-Escuredo and del Alamo, 2020).

Closely related to this, we find the issue about the neutrality of algorithms. As algorithms are designed by people to perform a certain task or respond to a specific question, algorithms convey potential design biases. Socio-economic interests are also at the core of AI development. Integrative visions of humans in the digital era should guide digitalization and AI development before human limitations and economic interests do. These visions should be built through interdisciplinary collaboration and should drive technical development instead of being reactive. Furthermore, education should engage technologists and ethics practitioners to raise the citizens, designers and builders of the digital era.

Bias identification and mitigation are a commonly known challenge of AI. Many biases are due to the limitations of the available data for learning. Generating better data is a mission that involves the radical collaboration of different stakeholders (public, private, academy, non-profit, etc) and ethical principles should be developed for a next generation of data-driven systems.

Monitoring AI performance and impact is key (Rahwan et al., 2019). Algorithms may work very differently depending on the context of use. The impact also greatly depends on the cultural context. From organizations to country and global level monitoring must be implemented with ethical principles that ensure rights, individuality, social good and cultural heterogeneity. An ethical framework to evaluate impact and deep implications of digitalization could guide regulation in terms of soft and hard law that ensure technological progress along with a responsible and sustainable evolution.

## 4 PERSPECTIVES

Machines will keep progressing. The ambition of AI research is to make machines perform human-like tasks, humanize matter even to the level of reductive physicalism. Emergence and supervenience suggest that machine will hardly reproduce human reasoning, but they can be very effective in performing human-like tasks. Some of these tasks are now possible with accuracy, speed and scalability. For instance, text analysis, translation and generation, and even more creative tasks such as painting or music composition are being achieved with surprising results (LeCun et al., 2015). We can expect that cultures will adopt and develop different levels and applications of AI because of their tradition, religion, industry and research system. We should start wondering about the social status of machines and how they will collaborate with humans to potentially take humanity to a next level of self-development and well-being.

New ethic needs to have a scientific ground. It is not sufficient that ethic reflects about science as a subject, science has to be part of ethic because it can help it to understand this complex world. As an alternative to negative techno-ethics that makes criticism about technology, it is necessary to promote a constructive ethic that creates an imaginary capable to drive technological development towards a better humanity. This landscape prompts for an exciting return to make questions about human conceptions and expectations by ethics.

## ACKNOWLEDGEMENTS

Author would like to thank the Center of Innovation and Technology for Development at Technical University Madrid and the Cátedra Iberdrola-UPM for support and collaboration. Also, thanks to the UN Global Pulse agency for collaboration.